\documentclass[prd,twocolumn,showpacs,floatfix,amsmath,nofootinbib,amssymb,floatfix]{revtex4}
\usepackage{graphicx,color,dcolumn,booktabs,bm}
\usepackage{longtable,lscape}
\usepackage{txfonts}
\usepackage{overpic}
\usepackage{amssymb}
\usepackage{indentfirst}
\usepackage{feynmf}   
\usepackage{slashed}  
\usepackage{cases}
\usepackage{color}
\usepackage{multirow}
\usepackage{slashed}
\usepackage{epstopdf}
\usepackage{longtable}
\usepackage{ulem}
\usepackage{graphicx,color,dcolumn,booktabs,bm}
\usepackage[colorlinks,
            citecolor=blue,
            anchorcolor=red,
            menucolor=red,
            linkcolor=red,
            filecolor=red,
            runcolor=red,
            urlcolor=blue,
            frenchlinks=red]{hyperref}

\graphicspath{{Figures/}} %
\allowdisplaybreaks

\begin{document}

\title{Studying $X(2100)$ hadronic decays and predicting its pion and kaon induced productions}

\author{Li-Ming Wang$^{1,2}$}\email{lmwang15@lzu.edu.cn}
\author{Jun-Zhang Wang$^{1,2}$}\email{wangjzh2012@lzu.edu.cn}
\author{Si-Qiang Luo$^{1,2}$}\email{luosq15@lzu.edu.cn}
\author{Jun He$^{3}$}\email{junhe@ninu.edu.cn}
\author{Xiang Liu$^{1,2}$\footnote{Corresponding author}}
\email{xiangliu@lzu.edu.cn}
\affiliation{
$^1$School of Physical Science and Technology, Lanzhou University, Lanzhou 730000, China\\
$^2$Research Center for Hadron and CSR Physics, Lanzhou University
and Institute of Modern Physics of CAS, Lanzhou 730000, China\\
$^3$Department of Physics and Institute of Theoretical Physics, Nanjing Normal University, Nanjing, Jiangsu 210097, China}

\begin{abstract}
The newly observed $X(2100)$ by the BESIII Collaboration inspires our interest in studying the light meson system, especially axial-vector mesons. Since the $X(2100)$ has $J^P=1^+$ possibilities but cannot be distinguished only by mass, we make use of flux-tube model
to study the strong decay behavior of $X(2100)$ under this assignment. The experimental width of the newly reported $X(2100)$ can be reproduced in our calculation, which favors an assignment of $X(2100)$ as the second radial excitation of $h_1(1380)$ with $I(J^P)=0(1^+)$. And the $\mathcal{B}(X(2100)\to \phi \eta^\prime)$ has a sizable contribution to the total width. Furthermore, we focus on the production of $X(2100)$ and its flavour partner $h_1(1965)$ induced by pion and kaon on a proton target with the Feynman model and the Regge model, which is an available platform to further identify their nature. The numerical results indicate that the total cross section are similar in the two models. When the range of momentum ${\mathrm{p_{Lab}}}$ is 10 to 30 GeV/$c$, the total cross sections for $\pi^-p\to X(2100)n$ and $K^-p\to X(2100)\Lambda$ are predicted to be at an order of magnitude of 0.1 $\mu$b.
Whereas, the total cross section for $\pi^-p\to h_1(1965)n$ is near an order of magnitude of 10 $\mu$b when $p_{\mathrm{Lab}}$ is from 10 to 30 GeV/$c$, and much larger than that of reaction $K^-p\to h_1(1965)\Lambda$. These predictions can provide some valuable information to search for $X(2100)$ and $h_1(1965)$ in experiments at J-PARC, COMPASS, OKA@U-70 and SPS@CERN.
\end{abstract}
\date{\today}
\maketitle

\section{Introduction}\label{sec1}

Since 2003, a series of light flavor $XYZ$ states have been observed, which include $X(1835)$ in $J/\psi\to\gamma \eta^\prime\pi^+\pi^-$ \cite{Ablikim:2005um,Ablikim:2010au}, $X(1860)$ in $J/\psi\to \gamma p\bar{p}$ \cite{Bai:2003sw}, $X(1812)$ in $J/\psi\to \gamma \omega\phi$ \cite{Ablikim:2006dw}, $Y(2175)$ in $J/\psi\to \eta\phi f_0(980)$ \cite{Ablikim:2007ab,Ablikim:2014pfc} and so on. These observations have stimulated theorists extensive interest in decoding their inner structures under either exotic state framework \cite{Li:2005bz,Ding:2005ew,Wang:2010vz,Li:2006ru,Chao:2006fq,Bicudo:2006sd,Klempt:2007cp,Wang:2006ri,Ding:2006ya,MartinezTorres:2008gy} or conventional meson system \cite{Wang:2017iai,Huang:2005bc,Yu:2011ta,Liu:2010tr,Li:2008mza,Wang:2012wa,Ding:2007pc}. These observed light flavor $XYZ$ states enlarged our knowledge of establishing light meson spectrum and exploring exotic hadronic states.

The progresses of exploring light flavor $XYZ$ states never stop. Recently, the BESIII Collaboration announced the observation of a structure in the $\phi \eta^{\prime}$ invariant mass spectrum of the $J/\psi \to \phi\eta\eta^{\prime}$ decay \cite{null}. With two assumptions of the spin-parity quantum number ($J^P=1^-$ and $1^+$), its resonance parameters are extracted, i.e.,
\begin{equation}
J^P=1^-:\left\{
\begin{split}
     M=&(2002.1\pm27.5\pm15.0)\,{\text{MeV}}\\
\Gamma=&(129\pm17\pm7)\,{\text{MeV}}
\end{split}\right. ,
\end{equation}
and
\begin{equation}
J^P=1^+:\left\{
\begin{split}
     M=&(2062.8\pm13.1\pm4.2)\,{\text{MeV}}\\
\Gamma=&(177\pm36\pm20)\,{\text{MeV}}
\end{split}\right. .\label{h1}
\end{equation}
However, the present angular distribution analysis cannot distinguish the $1^-$ and $1^+$ assumptions due to the limited statistics \cite{null}.

Before the BESIII observation, Lanzhou group once systematically studied axial-vector meson family by analyzing mass spectrum and calculating two-body Okubo-Zweig-Iizuka (OZI)-allowed decays \cite{Chen:2015iqa}. Here, as the second radial excitation of $h_1(1380)$, $h_1(2120)$ was predicted, whose resonance parameter is consistent with the BESIII's measurement listed in Eq. (\ref{h1}). This fact makes us conjecture that this structure existing in the $\phi \eta^{\prime}$ invariant mass spectrum of the $J/\psi \to \phi\eta\eta^{\prime}$ decay \cite{null} should be a good candidate of higher state in the $h_1$ meson family. In the following discussion, this structure reported in the $J/\psi \to \phi\eta\eta^{\prime}$ decay \cite{null}  is named as $X(2100)$ tentatively. We need to emphasize that both $J/\psi\to X(2100) \eta$ and $X(2100)\to \phi \eta^{\prime}$ are typical $S$-wave decays under the axial-vector meson assignment to $X(2100)$.

In this work, we briefly illustrate the reasonability of categorizing $X(2100)$ as the second radial excitation of $h_1(1380)$.
Comparing with former work \cite{Chen:2015iqa}, the present work provide more comprehensive information of two-body OZI-allowed decay behavior of $X(2100)$, where we adopt the flux-tube model \cite{Wang:2017iai,Isgur:1984bm,Deng:2012wj,Kokoski:1985is,Blundell:1996as} to deal with the calculation of strong decays. It is obvious that the information of decay can be applied to further study relevant to $X(2100)$. Concretely, we can extract the branching fraction of $J/\psi\to X(2100)\eta$ by
the product branching fraction $\mathcal{B}(J/\psi\to \eta X(2100))\times \mathcal{B}(X(2100)\to \phi \eta^\prime)=(9.6\pm1.4\pm1.6)\times 10^{-5}$, which was given by BESIII \cite{null}.
And then, the product branching fraction $\mathcal{B}(J/\psi\to \eta X(2100))\times \mathcal{B}(X(2100)\to \phi \eta)$ can be estimated. According to it, we may expect that experimental search for $X(2100)$ becomes possible via $J/\psi\to \eta\eta\phi$. The detailed illustration can be found in the following sections.

Furthermore, the present work will focus on the production of $X(2100)$ via pion and kaon induced reactions.
In the concrete calculation, we will adopt effective Lagrangian approach. The obtained partial decay widths of $X(2100)$ can be applied to extract the corresponding coupling constants, which will be input parameters when calculating the cross section of these discussed production processes. Since $X(2100)$ only was observed in a hadronic decay of $J/\psi$ \cite{null}, searching for $X(2100)$ via other processes will be interesting issue. The present study can provides valuable information for further experiments.
This paper is organized as follows. After introduction, the categorization of $X(2100)$ in $h_1$ meson family is performed in Sec. \ref{sec2}. And then, in Sec. \ref{sec3}, we explore the production of $X(2100)$ induced by pion or kaon. The paper ends with a short summary in Sec. \ref{sec4}.

\section{Categorizing $X(2100)$ in $h_1$ meson family}\label{sec2}


In this section, we perform a study of the spectrum and OZI-allowed strong decay behavior of the newly observed $X(2100)$ when trying to categorize it in the corresponding meson family.

{The Regge trajectories assume some mass dependence for states with different total spin, which satisfies the relation $M_J^2=M_{J^\prime}^2+\alpha^2(J-J^\prime)$, where where $J$ or $J^\prime$ is the spin of a meson, $M_{J}$ and $M_{J^\prime}$ denote the masses of mesons with different spins and with the same $P$ and $C$ quantum numbers. The discussed $h_1$ states with $I(J^{PC})=0(1^{+-})$ can construct Regge trajectories by combing with $h_3$ states with $0(3^{+-})$ and $h_5$ states with $0(5^{+-})$. When checking Particle Data Group (PDG) data \cite{Tanabashi:2018oca}, we notice that there only exist two $h_{3}$ states ($h_3(2025)$ and $h_3(2275)$) and $h_5$ states are still missing in experiment, which make the difficulty of quantitatively estimating the masses of higher $h_1$ states by $J-M_{J}^2$ plot. Considering this status, in this work we adopt another analysis of Regge trajectories, which is a non-standard way. }

Since {this} approach of Regge trajectory analysis \cite{Chew:1962eu,Kovalenko:2015ssa} was extensively adopted and tested in different meson systems (see Refs. \cite{,Ebert:2009ub,Yu:2011ta,Wang:2017iai,Anisovich:2000kxa,Chen:2015iqa,Ye:2012gu,He:2013ttg,Guo:2019wpx} for more details), in this work we plot Regge trajectories to analyze $X(2100)$ with formula
\begin{eqnarray}\label{eqRegge}
M^2=M_0^2+(n-1)\mu^2,
\end{eqnarray}
where $M_0$ is the ground state mass and $M$ denotes the mass of radial excitation with radial quantum number $n$. $\mu^2$ is the slope parameter of the trajectory \cite{Anisovich:2000kxa}.

\begin{figure}[h]
\centering
\begin{tabular}{cc}
\includegraphics[width=0.43\textwidth,scale=0.4]{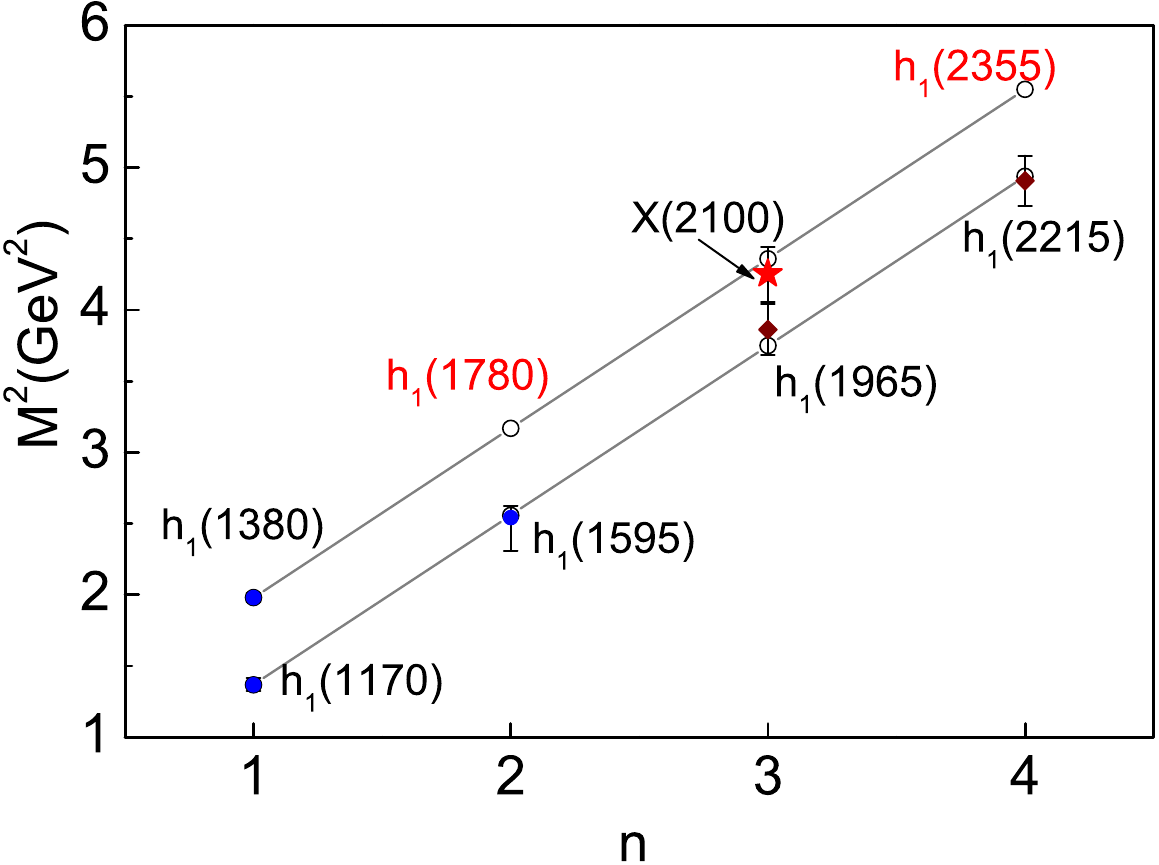}
\end{tabular}
\caption{(Color online). The analysis of the Regge trajectories for these $h_1$ states with $\mu^2=1.19$ GeV$^2$. Here, the solid circles with blue denote experimental values. Hollow circles are theoretical values and wine rhombuses denote further states listed in PDG \cite{Tanabashi:2018oca}. The newly observed $X(2100)$ under the $J^P=1^+$ assignment is marked by the red star.}\label{figRegge}
\end{figure}

If fixing slope parameter $\mu^2=1.19$ GeV$^2$ and taking the mass of $h_1(1170)$ and $h_1(1380)$ as input, we can construct Regge trajectory, and find that the mass of the second radial excitation of $h_1(1380)$ is 2087 MeV, which overlaps with the measured mass of $X(2100)$ by BESIII with the $J^P=1^+$ assumption \cite{null}. The analysis of Regge trajectories for these discussed $h_1$ states {is} shown in Fig. \ref{figRegge}. It shows that assigning $X(2100)$ into the isoscalar axial-vector meson family is possible.

In following discussion, we focus on OZI-allowed strong decay behavior of the newly observed $X(2100)$ as an axial-vector in the flux tube model \cite{Wang:2017iai,Isgur:1984bm,Deng:2012wj,Kokoski:1985is,Blundell:1996as}, by which the reliability of this assignment can be tested further. The model is applied to quantitatively give the decay information of hadron. In the flux-tube model, a quark and antiquark connected by a tube of chromoelectric
flux construct a meson, where this tube can be treated as a vibrating string. For the meson, the string is in the vibrational ground state (vibrational excitation corresponds to hybrid). When a meson decay occurs, the vibrational string breaks at a point, and simultaneously quark-antiquark pair is created from vacuum to connect to the free end of string and further form two outgoing mesons. The details of the flux-tube model can be refereed to Refs. \cite{Isgur:1984bm,Blundell:1996as}.

Before experimental observation, Lanzhou group have systematically studied the decay behavior of axial-vector mesons \cite{Chen:2015iqa} in $^3P_0$ model. However, the decay channel $\phi\eta^{\prime}$
is not considered in their calculation which is an important channel because of observation of $X(2100)$ in this constant mass spectrum. Therefore, we calculate the strong
decay behavior of $X(2100)$ under axial vector meson assignment again but in flux tube model.  As is known, the second radial excitation of $h_1(1380)$ is mainly
composed of $s\bar{s}$, and $X(2100)$ is observed in the $\phi\eta^{\prime}$ invariant mass spectrum, so $X(2100)$ is a good candidate of $h_1(1380)$'s radial excitation.

\begin{figure}[hbtp]
\centering
\includegraphics[width=0.27\textwidth,scale=0.4]{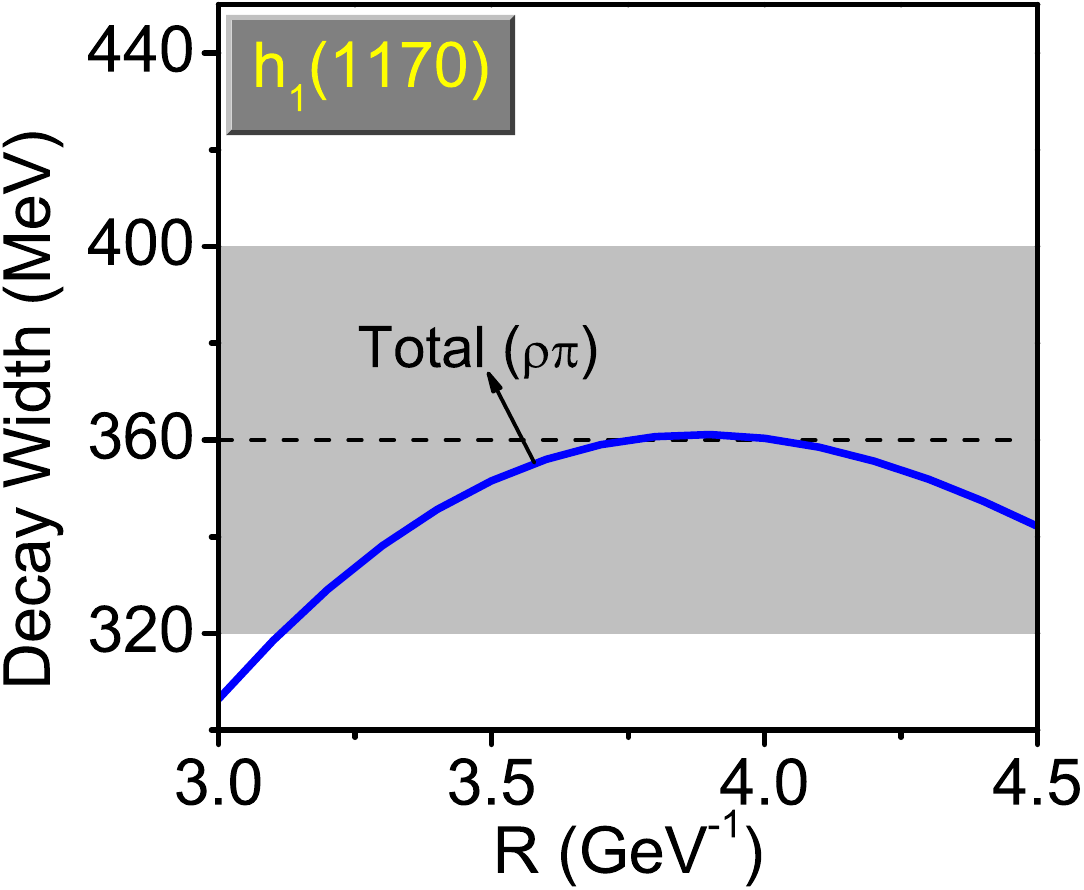}\\
\includegraphics[width=0.47\textwidth,scale=0.4]{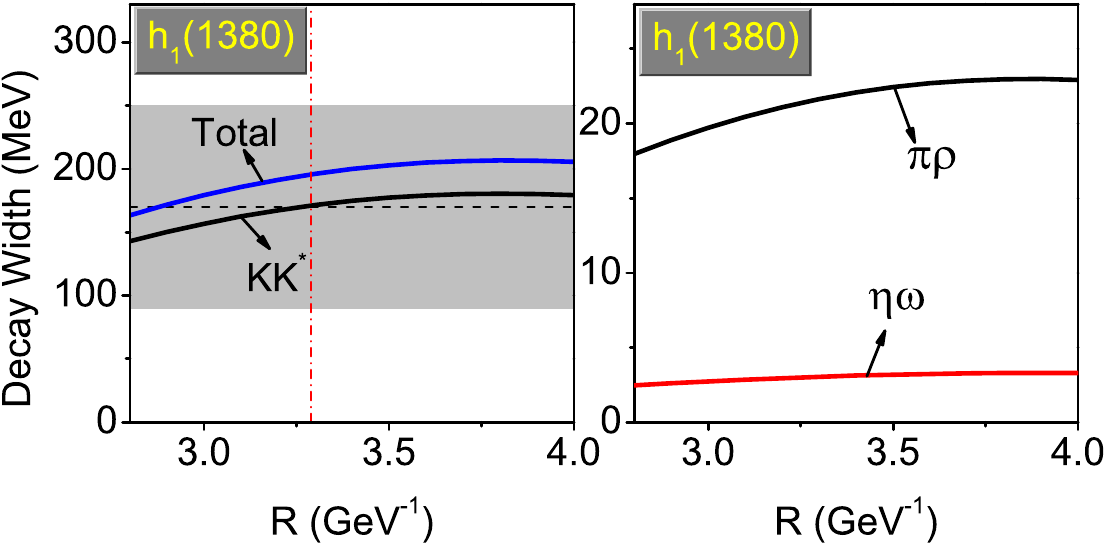}
\caption{(Color online). The $R$ dependence of the total and partial decay widths of $h_1(1170)$ \cite{Dankowych:1981ks,Ando:1990ti,Atkinson:1983yx} and $h_1(1380)$ \cite{Aston:1987ak,Abele:1997vu,Ablikim:2015lnn,Ablikim:2018ctf}, where listed the corresponding experimental data (dashed lines with gray band) for comparison with our theoretical calculation. The selected $R$ range already contains the suggested $R$ value for ground state of the $h_1$ family in Ref. \cite{Close:2005se}.}\label{figh11170h11380}
\end{figure}

In order to test the realiability of the model and these adopted parameters , we study the ground states of $h_1$ family. Usually,
a mixing scheme should be introduced when discussing $h_1(1170)$ and $h_1(1380)$, i.e.,
\begin{equation}\label{mixingangleh11P}
\left( \begin{matrix}
	|h_1(1170)\rangle \\
	|h_1(1380)\rangle \\
\end{matrix}\right) =
\left( \begin{matrix}
	\textrm{$\sin\theta_1$} & \textrm{$\cos\theta_1$} \\
	\textrm{$\cos\theta_1$} & \textrm{$-\sin\theta_1$} \\
\end{matrix}\right)
\left( \begin{matrix}
	|n\bar{n}\rangle \\
	|s\bar{s}\rangle \\
\end{matrix}\right),
\end{equation}
where $n\bar{n}=(u\bar{u}+d\bar{d})/\sqrt{2}$. $\theta_1$ is the mixing angle. The concrete value of $\theta_1$ is suggested to be $\theta_1\sim82.7^\circ$ \cite{Cheng:2011pb}, $\theta_1=86.8^\circ$ \cite{Dudek:2011tt}, $\theta_1=85.6^\circ$ \cite{Li:2005eq} by different theoretical groups. In our calculation, we adopted $\theta_1=78.7^\circ$ from the estimate by Gell-Mann-Okubo formula. And then, the obtained decay behavior is shown in Fig. \ref{figh11170h11380}. We can see that $h_1(1170)$ and $h_1(1380)$ are well established ground states in the $h_1$ meson family.

\begin{figure}[hbtp]
\centering
\includegraphics[width=0.45\textwidth,scale=0.5]{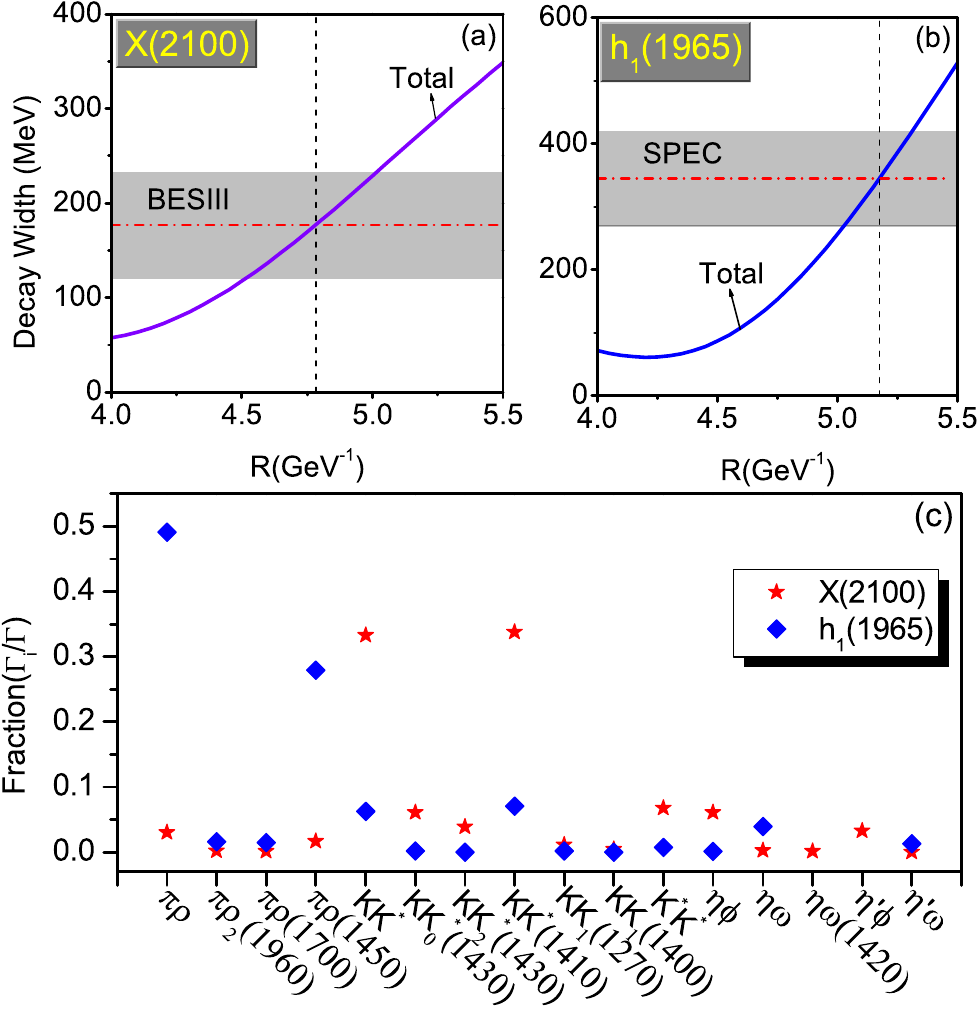}
\caption{(Color online). The strong decay behaviors of $X(2100)$ and $h_1(1965)$, which are signed as the second excited state of the axial meson. (a) and (b) are the total decay width of $X(2100)$ and $h_1(1965)$, respectively in comparison with the experimental data (dashed lines with gray band). (c) The branch ratio of decay channels of $X(2100)$ (red star) and $h_1(1965)$ (blue diamond) with $R$=4.78 GeV$^{-1}$ and 5.17 GeV$^{-1}$, respectively.}\label{br2119}
\end{figure}

The key point of the present work is to test the possibility of the newly observed $X(2100)$ as the second radial excitation of the $h_1$ family, which becomes a partner of $h_1(1965)$. $X(2100)$ and $h_1(1965)$ satisfy the below relation
\begin{equation}\label{mixingangleh13P}
\left( \begin{matrix}
	|h_1(1965)\rangle \\
	|X(2100)\rangle \\
\end{matrix}\right) =
\left( \begin{matrix}
	\textrm{$\sin\theta_2$} & \textrm{$\cos\theta_2$} \\
	\textrm{$\cos\theta_2$} & \textrm{$-\sin\theta_2$} \\
\end{matrix}\right)
\left( \begin{matrix}
	|n\bar{n}\rangle \\
	|s\bar{s}\rangle \\
\end{matrix}\right).
\end{equation}
The total width of $X(2100)$ under $h_1(1380)$'s second radial excitation assignment is shown in Fig. \ref{br2119} when taking $\theta=78.7^{\circ}$ \footnote{We can get the mixing angle from Gell-Mann-Okubo formula, using the masses of ground states.
Generally, the $\theta=$arctan$\frac{m_{18}^2}{m_{h_1^\prime}^2-m_8^2}+$arctan$\sqrt{2}$ is the same as ground states if the masses of excited nonet obtained from Regge trajectory.}

We can see that the experimental width of $X(2100)$ can be reproduced well, which shows that $X(2100)$ can be {a} good candidate of the second radial excitation of the $h_1$ meson family. The corresponding branching ratios are given in Fig. \ref{br2119}. Here, $KK^*(1410)$ and $KK^*$ as main decay channels and $K^*K^*$, $KK_0^*(1430)$, $\eta\phi$, $KK_2^*(1430)$ $\eta^\prime\phi$, $\pi\rho$ and $\pi\rho(1450)$ as subordinate decays can be identified.
But, there exist some challenges to identify kaon since weak interaction dominates the decay of kaon, which makes the whole reconstruction efficiency of $X(2100)$ via these kaon final states
is low. It may be the reason why $X(2100)$ cannot be identified firstly {from} the kaon final states.

Since $X(2100)\to\phi\eta^{\prime}$ has a sizable contribution to the total width, $X(2100)$ firstly observed in the $\phi\eta^{\prime}$ final state can be understood.
Our calculation gives that the branching ratio of $X(2100)\to\phi\eta^{\prime}$ can reach up to $3.29\%$ as seen in Fig. \ref{br2119}. Furthermore, branching ratio
does not change much with changes in $R$. The BESIII Collaboration measured the product branching
fraction $\mathcal{B}(J/\psi\to \eta X(2100))\times \mathcal{B}(X(2100)\to \phi \eta^\prime)=(9.6\pm1.4\pm1.6)\times 10^{-5}$. Combining our theoretical result and this
experimental data, we may extract
$$\mathcal{B}(J/\psi\to \eta X(2100))=2.92\times10^{-3}.$$
This branching ratio provides important information of further theoretical study on the production of $X(2100)$ associated with a $\eta$ meson via the $J/\psi$ decay.

\begin{figure}[hbtp]
\centering
\end{figure}

Additionally, we notice that $X(2100)\to\phi\eta$ with branching ratio $6.06\%$ is comparable with $X(2100)\to\phi\eta^{\prime}$.  It is obvious that this product branching
fraction is comparable with $\mathcal{B}(J/\psi\to \eta X(2100))\times \mathcal{B}(X(2100)\to \phi \eta^\prime)$.
Thus, experimental search for $X(2100)$ via $J/\psi\to \phi\eta\eta$ is suggested, which can be as a potential experimental issue at BESIII.

As the partner of $X(2100)$, the decay {behaviors} of $h_1(1965)$ can be obtained, which is illustrated in Fig. \ref{br2119}. $h_1(1965)$ that is listed in PDG as a further state has not been determined. If reproducing the central value of the width of $h_1(1965)$, we should take $R=5.17$ GeV$^{-1}$. The corresponding typical decay channels are $\rho\pi$, $\rho(1450)\pi$, which almost determine the total width of $h_1(1965)$. Then, $\eta\omega$, $KK^*(1410)$ and $KK^*$ have sizable contribution to the total decay width, which may explain why the {Crystal Barrel} experiment announced the observation of  $h_1(1965)$ in  the $p\bar{p}\to \omega \eta$ process \cite{Anisovich:2011sva}.



In the near future, BESIII will play {an} important role to explore light hadrons. Our results also show that $h_1(1965)$ may decay into $\eta\omega$ which has comparable branching fraction with that of $\eta^\prime\omega$ (see Fig. \ref{br2119}), i.e.,
\begin{eqnarray}\label{equa13}
\frac{\Gamma(h_1(1965)\to\omega\eta)}{\Gamma(h_1(1965)\to\omega\eta^\prime)}=3.0
\end{eqnarray}
 We suggest BESIII to carry out the study of $J/\psi\to \eta h_1(1965)\to\eta \eta^{\prime} \omega$ and $\eta\eta\omega$, especially by checking the corresponding $\omega\eta$ and $\omega\eta^\prime$ invariant mass spectra. Since $h_1(1965)$ is a broad structure, how to identify this broad resonance will be a challenge for experimentalist.

\section{Exploring the production of $X(2100)$}\label{sec3}
From the above analysis of mass spectrum and hadronic decay behavior, it is appropriate to categorize $X(2100)$ into axial-vector $h_1$ family. However, we still need more experimental data to further confirm the assignment of $X(2100)$ as $h_1(3P)$. Since $X(2100)$ is only observed in the $J/\psi$ hadronic decay, so its experimental search in different reaction platforms will become important. In the past four decades, the reaction induced by pion or kaon on a proton target is a very important kind of experimental tool to search for light hadrons, especially the axial-vector $h_1$ mesons. For example, both $h_1(1170)$ \cite{Dankowych:1981ks,Ando:1990ti} and $h_1(1595)$ \cite{Eugenio:2000rf} are produced in the early experiments via pion induced reaction on a proton target. Similarly, as the partner of the ground state of $h_1(1170)$, $h_1(1380)$ also has been obtained from reaction $K^-p\to h_1(1380)\Lambda$ \cite{Aston:1987ak} in the LASS experiment. Therefore, pion-proton and kaon-proton scattering could be alternative methods to study the properties of $X(2100)$ and other axial-vector mesons. At the same time, the research results of hadronic decays in Sec. \ref{sec2} can be used to extract coupling constants of the interaction vertex, which can allow us to make a quantitative calculation of pion and kaon induced reaction.

Based on the above motivations, we will investigate the productions of $X(2100)$ and its partner $h_1(1965)$ via pion and kaon induced reaction on a proton target by an effective Lagrangian approach. These theoretical predictions are expected to be valuable for the relevant experimental measurements, which include J-PARC \cite{Kumano:2015gna} and COMPASS \cite{Nerling:2012er} with pion beam, and OKA@U-70 \cite{Obraztsov:2016lhp}, SPS@CERN \cite{Velghe:2016jjw} and J-PARC \cite{Nagae:2008zz} with kaon beam.

\begin{figure}[hbtp]
\centering
\includegraphics[width=0.5\textwidth,scale=0.4]{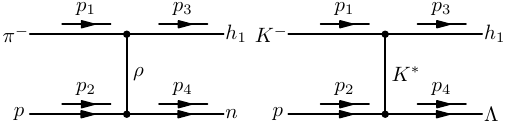}
\caption{The Feynman diagrams for the productions induced by pion and kaon. The left is $\pi^-p\to h_1n$ reaction and the right is $K^-p\to h_1\Lambda$ reaction.}\label{lag}
\end{figure}

The Feynman diagrams of induced reactions of pion and kaon on a proton target are illustrated in Fig. \ref{lag}, where only $t$-channel diagram is considered.
In general, the contribution from the $s$-channel is expected to be very small here because the mass of intermediate nucleon pole is much smaller than the production threshold of $X(2100)$ or $h_1(1965)$. In addition, since the decay width of $X(2100)/h_1(1965)\to p\bar{p}$ is not dominant, the $u$-channel contribution is usually negligible at low energies. At higher energies, the $u-$channel contribution always concentrates at backward angles, which can be easily distinguished from $t-$channel contribution at forward angles. Thus, in the present work, we only consider the contribution of vector meson exchange in $t$-channel and ignore the contributions from the nucleon resonances in the $s$- and $u$-channels.
As seen in Fig. \ref{br2119}, the $\pi\rho$ and $KK^*$ are main two-body decay channels of $h_1(1965)$ and $X(2100)$. Hence, the only $\rho$ and $K^*$ exchange are considered.

For the reaction of $\pi^-p\to h_1n$, the related effective Lagrangian that is shown in the left of Fig. \ref{lag} can be written as \cite{Kochelev:2009xz,Domokos:2009cq,Colangelo:2010te,Liu:2008qx}
\begin{eqnarray}\label{L1}
\begin{split}
\mathcal{L}_{h_1\rho\pi}=&g_{h_1\rho\pi}\ h_1^{\mu}\ \vec{\rho}_{\mu}\cdot\vec{\pi},\\
\mathcal{L}_{\rho NN}=&-g_{\rho NN}\ \bar{N}(\gamma^{\mu}-\frac{\kappa_{\rho}}{2m_N}\cdot\sigma^{\mu\nu}\partial_{\nu})\vec{\tau}\cdot\vec{\rho}_{\mu}\ N,
\end{split}
\end{eqnarray}
where $N, h_1, \rho$ and $\pi$ denote the nucleon, $X(2100)$ or $h_1(1965)$, $\rho(770)$ and pion field, respectively. Here, the coupling constants $g_{\rho NN}=3.36$ and $\kappa_{\rho}=6.1$ are adopted as suggested in Refs. \cite{Kochelev:2009xz,Domokos:2009cq}. Besides, the coupling constants $g_{X(2100)\rho\pi}$ and $g_{h_1(1965)\rho\pi}$ can be determined by the calculated decay width of $X(2100) \to \rho\pi$ and $h_1(1965) \to \rho\pi$ in the flux tube model, respectively. The BESIII data \cite{null} indicate that the central total width of $X(2100)$ is 177 MeV under the $J^P=1^+$ assumption, which corresponds to a partial width of 5.5 MeV for the $X(2100)\to\rho\pi$ in our theoretical calculation. In the effective Lagrangian framework, this width will lead to a value of coupling constant $g_{X(2100)\rho\pi}=0.395$. Similarly, the central total width of $h_1(1965)$ as a further state in PDG is suggested to be 345 MeV \cite{Anisovich:2011sva} and so partial width of $h_1(1965)\to\rho\pi$ we calculated is 177.34 MeV, from which the coupling constant $g_{h_1(1965)\rho\pi}=1.83$ can be obtained.

For the reaction of $K^-p\to h_1\Lambda$, the related effective Lagrangian that is shown in Fig. \ref{lag} can be written as \cite{Colangelo:2010te,Wan:2015gsl,Wang:2015xwa}
\begin{eqnarray}\label{L2}
\begin{split}
\mathcal{L}_{h_1KK^*}=&g_{h_1KK^*}\ h_1^{\mu}\ (K_{\mu}^{*\dag}K+K^{\dag}K_{\mu}^*),\\
\mathcal{L}_{K^*N\Lambda}=&-g_{K^*N\Lambda}\ \bar{N}(\slashed{K}^*-\frac{\kappa_{K^*N\Lambda}}{2m_N}\sigma_{\mu\nu}\partial^{\nu}K^{*\nu})\Lambda+\mathrm{H.c}.,
\end{split}
\end{eqnarray}
where $K, K^*$ and $\Lambda$ denote kaon, $K^*$ and hyperon $\Lambda$ field, respectively, and $m_N$ is the mass of nucleon. By the same way, we can get the coupling constants $g_{X(2100)KK^*}=1.32$ and $g_{h_1(1965)KK^*}=0.816$ from the decay behaviors of $X(2100) \to KK^*$ and $h_1(1965) \to KK^*$. The coupling constants $g_{K^*N\Lambda}=-4.26$ and $\kappa_{K^*N\Lambda}=2.66$ were given by the Nijmegen potential \cite{Stoks:1999bz}.

For the $t$-channel exchange, we introduce a monopole type form factor $F_{v}(k)=(\Lambda_{v}^2-m_{v}^2)/(\Lambda_{v}^2-k^2)$ at each interaction vertex \cite{Machleidt:1987hj,Machleidt:1989tm}, where $k$ and $m_{v}$ are the four-momentum and mass of the exchanged meson, respectively, where subscript $v$ represents the corresponding exchanged vector meson, such as $\rho$ and $K^*$. The main reason of introducing form factor is to compensate the off-shell effects of the exchanged meson and depict the inner structure of every interaction vertex at the hadron level \cite{Tornqvist:1993ng,Tornqvist:1993vu}. The value of the cutoff $\Lambda_{v}$ can be parameterized to be $\Lambda_{v}=m_{v}+\alpha\Lambda_{\mathrm{QCD}}$, where $\Lambda_{\mathrm{QCD}}=220$ MeV \cite{Liu:2006dq}. In general, the cutoff parameter $\alpha$ is taken near 1.0 \cite{Cheng:2004ru}, so we take $\alpha$=1.0 in this work when giving the concrete numerical results.

According to the above effective Lagrangians, the scattering amplitude of the $\pi^-p\to h_1n$ can be obtained
\begin{eqnarray}\label{L2q}
\mathcal{M}_{\pi^-p\to h_1n}=&-g_{\rho NN}\ g_{h_1\rho\pi}\sqrt{2}\ \bar{u}(p_4)\left(\gamma^{\mu}+\frac{\kappa_{\rho}}{m_N}\sigma^{\mu\nu}(p_{t\nu})\right)\nonumber\\
&\times u(p_2)\ \frac{\tilde{g}^{\alpha\mu}}{p_t^2-m_{\rho}^2}\ \epsilon_{h_1}^{*\alpha}\ F_{\rho}^2(p_t),
\end{eqnarray}
where $p_2$ and $p_4$ are the four-momentum of incident proton and outgoing $h_1$, respectively (see Fig. \ref{lag}), and $p_t=p_4-p_2$ describes the four-momentum of exchange meson $\rho$. $\epsilon_{h_1}^{*\alpha}$ is the polarization vector of $h_1$ meson, and $\tilde{g}^{\alpha\mu}=-g^{\alpha\mu}+\frac{p_t^{\mu}p_t^{\alpha}}{m_{\rho}^2}$. The $u(p_2)$ and $\bar{u}(p_4)$ are the Dirac spinor of proton and neutron, respectively. For the process $K^-p\to h_1\Lambda$, the scattering amplitude reads as
\begin{eqnarray}\label{L2w}
\mathcal{M}_{K^-p\to h_1\Lambda}=&-g_{K^*N\Lambda}g_{h_1KK^*}\bar{u}(p_4)\left(\gamma^{\mu}-\frac{\kappa_{K^*N\Lambda}}{2m_N}\sigma_{\mu\nu}(-ip_t^{\nu})\right)\nonumber\\
&\times u(p_2)\frac{\tilde{g}^{\alpha\mu}}{p_t^2-m_{K^*}^2}\epsilon_{h_1}^{*\alpha}F_{K^*}^2(p_t),
\end{eqnarray}
where $\tilde{g}^{\alpha\mu}=-g^{\alpha\mu}+\frac{p_t^{\mu}p_t^{\alpha}}{m_{K^*}^2}$.

In order to better describe the behavior of the hadron production at high momentum, the Reggeized treatment should be introduced to the $t$-channel,  as well as the Feynman model above\cite{Guidal:1997hy,He:2010ii,Galata:2011bi,Haberzettl:2015exa,Wang:2015hfm,Levy:1973aq,Huang:2008nr}. The Regge model can be introduced by replacing the product of the form factor in Eq. (\ref{L2q}) and Eq. (\ref{L2w}) as {
\begin{eqnarray}\label{L2}
F_t^2(q_{\rho})\to\mathcal{F}_t^2(q_{\rho})&=&\left(\frac{s}{s_{\mathrm{scale}}}\right)^{\alpha_{\rho}(t)-1}
\frac{\pi\alpha'_{\rho}(t-m_{\rho}^2)}{\Gamma[\alpha_{\rho}(t)]\mathrm{sin}[\pi\alpha_{\rho}(t)]} \nonumber \\
&&\times (1+\xi e^{-i\pi\alpha_{\rho}(t)}),
\end{eqnarray}

\begin{eqnarray}\label{L2}
F_t^2(q_{K^*})\to\mathcal{F}_t^2(q_{K^*})&=&\left(\frac{s}{s_{\mathrm{scale}}}\right)^{\alpha_{K^*}(t)-1}
\frac{\pi\alpha'_{K^*}(t-m_{K^*}^2)}{\Gamma[\alpha_{K^*}(t)]\mathrm{sin}[\pi\alpha_{K^*}(t)]} \nonumber \\
&&\times (1+\xi e^{-i\pi\alpha_{K^*}(t)}).
\end{eqnarray}
The scale factor $s_{\mathrm{scale}}$ is fixed at 1 GeV, and $\xi$ is signature and $\xi=-1$ for the $\rho$-meson or $K^*$-meson exchange~\cite{Guidal:1997hy,Huang:2008nr}. It is worth mentioning that $e^{-i\pi\alpha_{\rho}(t)}=\mathrm{cos}[\pi\alpha_{\rho}(t)]-i\mathrm{sin}[\pi\alpha_{\rho}(t)]$ and the poles at positive $t$ include the $\rho(770)$ ($\alpha_{\rho}=1$), $\rho_3(1690)$ ($\alpha_{\rho}=3$) and so on, and the contributions of the poles with $\alpha_{\rho}=2,4,...$ can be avoided automatically. } In addition, the Regge trajectories of $\alpha_{\rho}$ and $\alpha_{K^*}$ read as \cite{Kochelev:2009xz,Ozaki:2009wp},
\begin{equation}\label{L2}
\begin{split}
\alpha_{\rho}(t)=&0.55+0.8t/\mathrm{GeV}^2,\\
\alpha_{K^*}(t)=&1+0.85(t-m_{K^*}^2)/\mathrm{GeV}^2.
\end{split}
\end{equation}
Here, as usual, we assume that only the low-mass first trajectory is dominate as suggested in Ref.~\cite{Corthals:2006nz}. For example, the daughter trajectory started from the $\rho(1450)$ is not included in the calculation.

With the the above preparations, the cross section of the $\pi^-p\to X(2100)n$, $K^-p\to X(2100)\Lambda$, $\pi^-p\to h_1(1965)n$ and $K^-p\to h_1(1965)\Lambda$ reactions will be calculated directly. The differential scattering cross section can be written as
\begin{eqnarray}\label{L2}
\begin{split}
\frac{\mathrm{d}\sigma}{\mathrm{d}t}=\frac{1}{64\pi s}\frac{1}{|p_{\mathrm{1cm}}|^2}|\mathcal{\overline{M}}|^2,
\end{split}
\end{eqnarray}
where $p_{\mathrm{1cm}}$ is the momentum of incident pion or kaon in the center of mass system and $t=(p_1-p_3)^2$ is square of the four momentum of the exchanged meson. $s=(p_1+p_3)^2$ is square of center of mass energy. The overline in $|\mathcal{\overline{M}}|^2$ means the average for the spin of initial states proton and the sum over the spin of final states $h_1$ and neutron(hyperon).

\begin{figure}[hbtp]
\centering
\includegraphics[width=0.45\textwidth,scale=0.4]{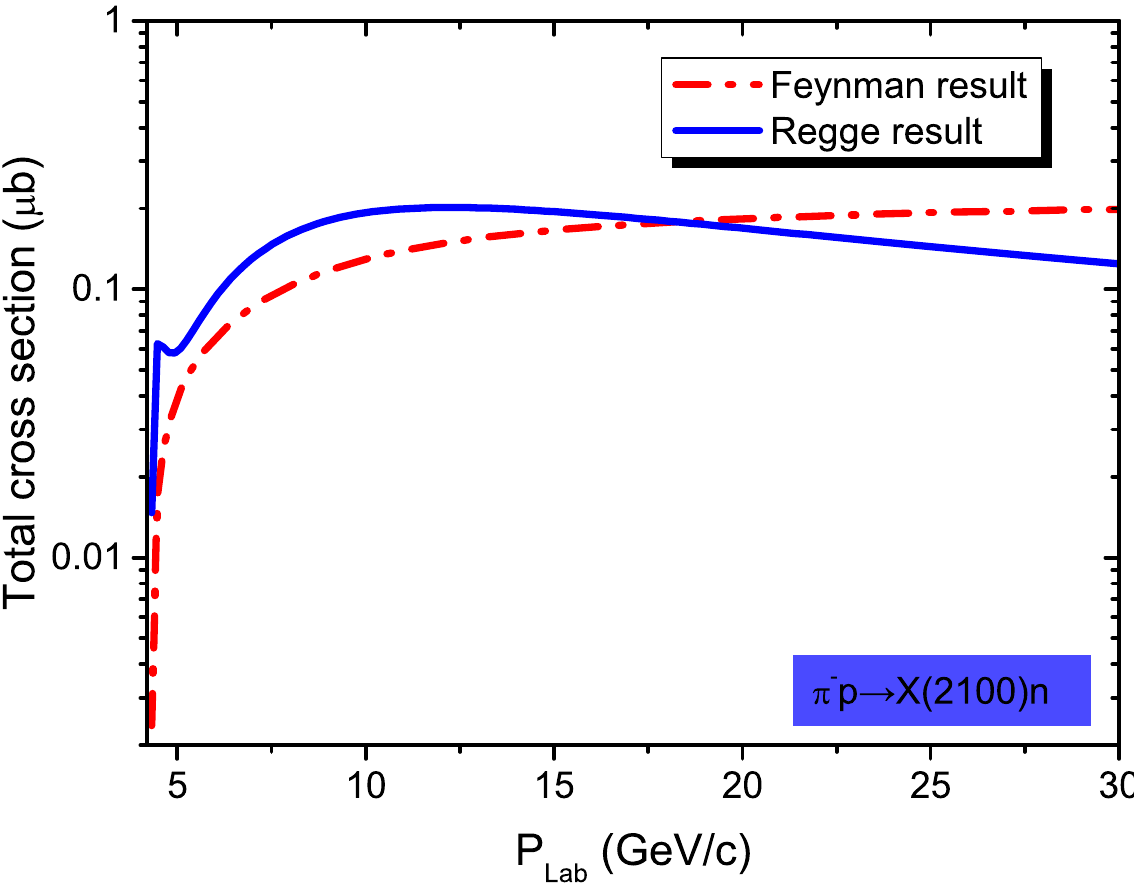}
\caption{Total cross section for the $\pi^-p\to X(2100)n$ reaction in Feynman model and Regge model.}\label{piNh1ss}
\end{figure}

From Fig.~\ref{piNh1ss}, we can see that the total cross section for $\pi^-p\to X(2100)n$ reaction is presented in the Feynman model and the Regge model. The line shape in both models are different. The line shape of total cross section in the Feynman model sharply increases near the threshold, but then slowly trend to a stable value. Whereas, in the Regge model, the total cross section increases sharply near the threshold and reaches a maximum of $0.20~\mu$b at a momentum ${\mathrm{p_{Lab}}}$ of 12.0 GeV$/c$. Different from the Feynman case, the line shape of total cross section in the Regge model decrease when ${\mathrm{p_{Lab}}}$ is greater than 12.0 GeV $/c$. We can also see from Fig. \ref{piNh1ss} that the Regge result is a little larger than that of the Feynman model within the momentum range from threshold to 18 GeV$/c$. However, as ${\mathrm{p_{Lab}}}$ continues to increase, the result of the Feynman model is greater. Overall, the total cross section for $\pi^-p\to X(2100)n$ are predicted to be at an order of magnitude of $0.1~\mu$b in both models. According to our theoretical predictions, the ${\mathrm{p_{Lab}}}$ range of 8.0 GeV$/c$ to 20 GeV$/c$ may be a suitable momentum window for future experiment to explore $X(2100)$ on the pion-proton scattering platform.

{Additionally, we notice that there exist a small cusp near the production threshold for the Regge result, it is relevant to the poles of $\Gamma[\alpha_\rho(t)]$ which can not be compensated by $\sin[\pi\alpha_\rho(t)]$ due to the existence of the factor $1-e^{-i\pi\alpha_{\rho}(t)}$, which leads to a dip at $\alpha_{\rho}(t)=0$, $-2$, $-4$ and so on. Such dip is also found in differential cross section of the $\pi N\to \pi N$ scattering~\cite{Huang:2008nr}. If we recall that the Regge model is more appropriate to describe the behavior at  high momentum and the near threshold behavior is more relevant to the Feynman exchange~\cite{Toki:2007ab, Nam:2010au,He:2012ud}, such cusp may be unphysical. The future precise experimental data is very helpful to clarify this issue.}

\begin{figure}[hbtp]
\centering
\includegraphics[width=0.45\textwidth,scale=0.4]{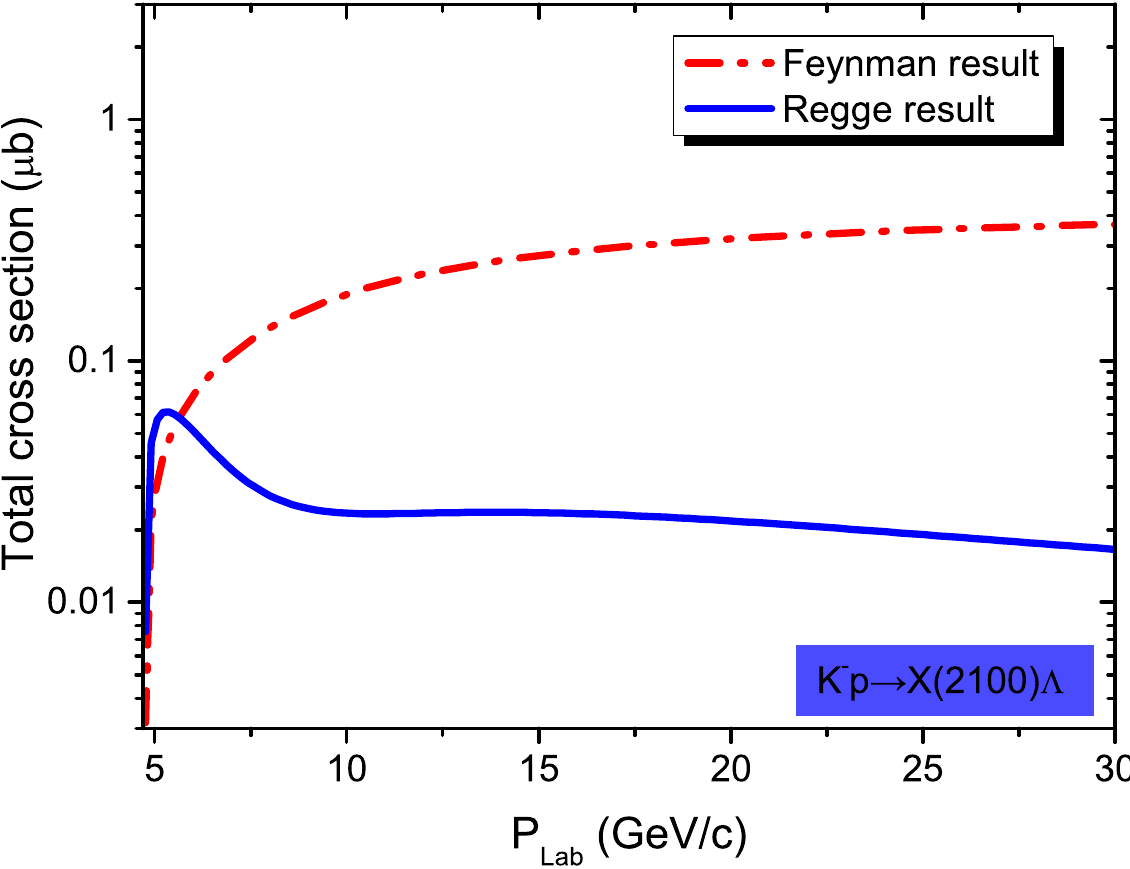}
\caption{Total cross section for the $K^-p\to X(2100)n$ reaction in Feynman model and Regge model.}\label{KNh1ss}
\end{figure}

In Fig. \ref{KNh1ss}, we present the total cross section of the $K^-p\to X(2100)\Lambda$ reaction within the Regge model and the Feynman model. Similar to the pion induced case, the line shape of total cross section in the Regge model is different from that in the Feynman model. In the Regge result, we can find that the line shape of total cross section goes up quiet rapidly and has a peak around ${\mathrm{p_{Lab}}}$=5.4 GeV$/c$. Compared to the Regge model, the Feynman model gives a larger cross section. The line shape difference between this two models is very helpful to clarify the role of the Regge model in further experiments.  Although ground state $h_1(1380)$ has been discovered in the kaon-proton scattering experiment \cite{Aston:1987ak}, but the experimental cross section data of kaon induced production of axial-vector $h_1$ states are still lacking. So we encourage experimentalist to pay attention to the production of excited $h_1$ states in both kaon-proton and pion-proton scattering, especially to $X(2100)$. It is very important to further understand their properties.


\begin{figure}[hbtp]
\centering
\includegraphics[width=0.45\textwidth,scale=0.4]{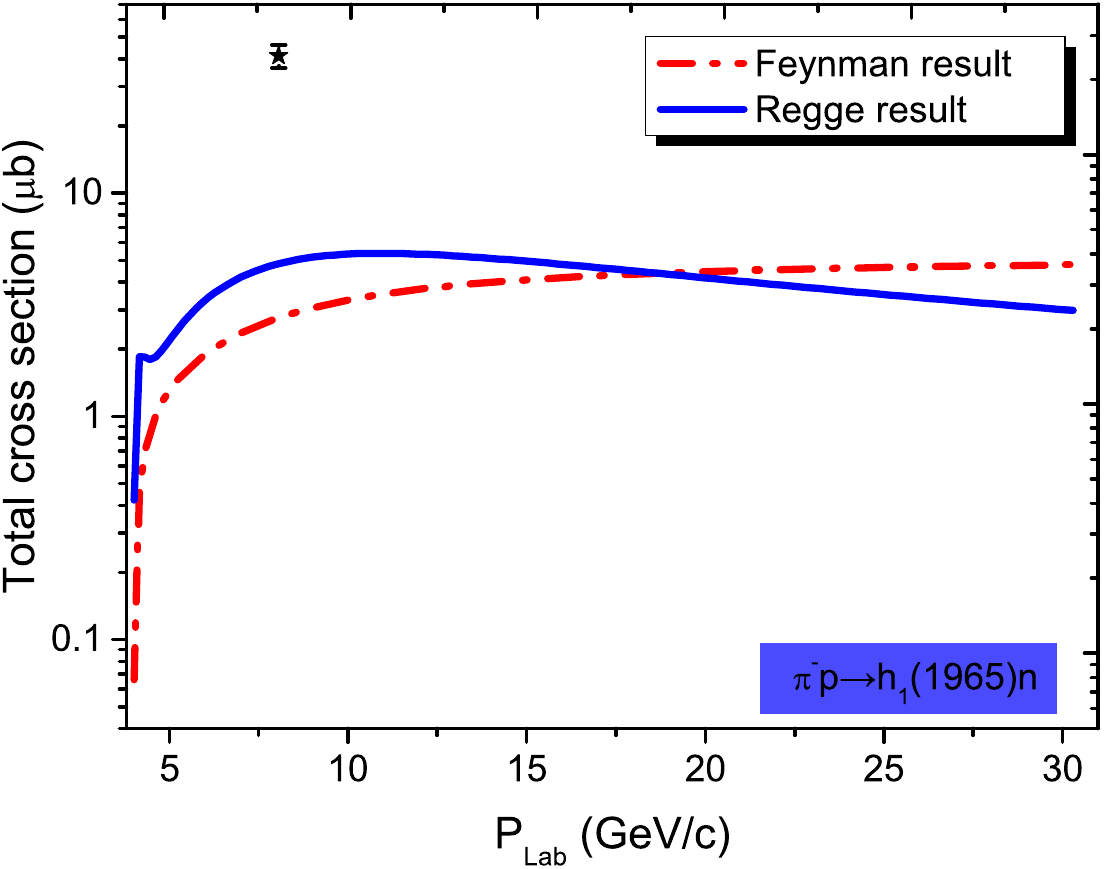}
\caption{Total cross section for the $\pi^-p\to h_1(1965)n$ reaction in Feynman model and Regge model. The point of black star with black bar is cross section of $\pi^-p\to h_1(1170)n$ reaction \cite{Ando:1990ti}.}\label{piNh1uu}
\end{figure}

\begin{figure}[hbtp]
\centering
\includegraphics[width=0.45\textwidth,scale=0.4]{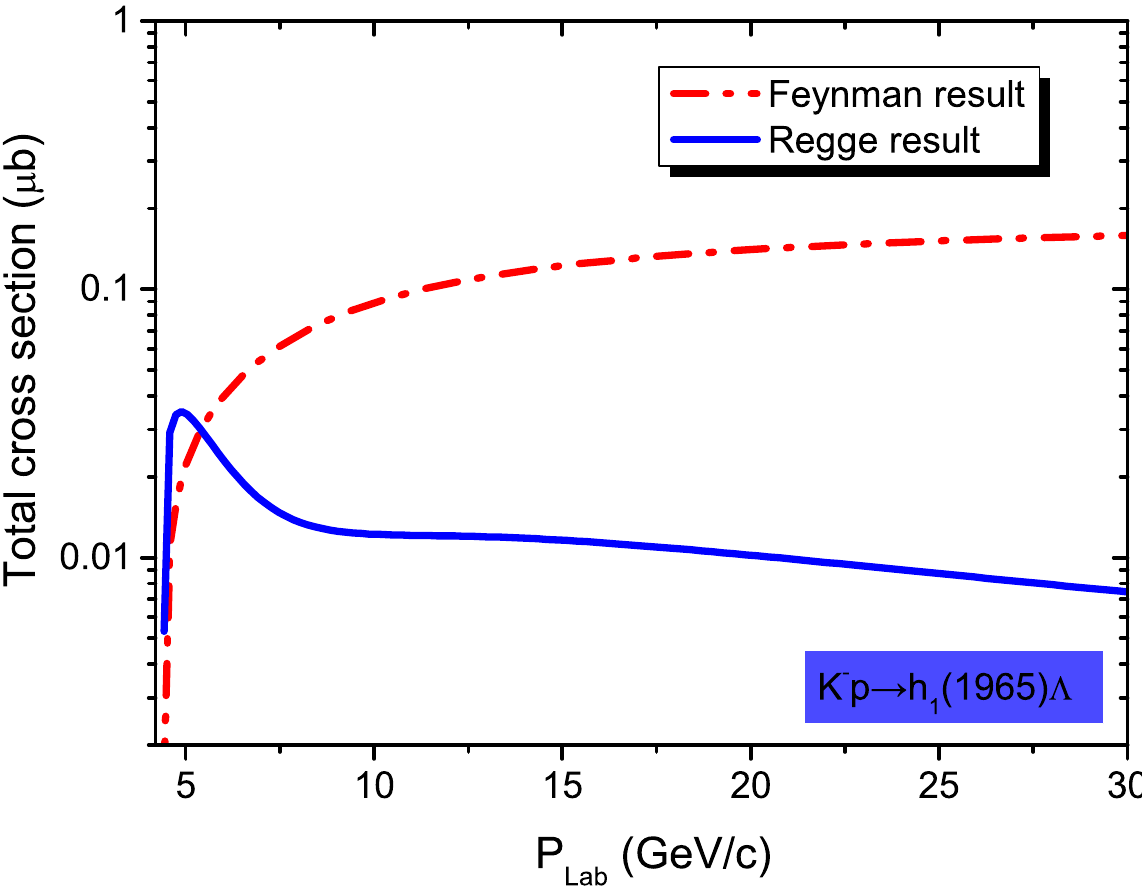}
\caption{The total cross section for the $K^-p\to h_1(1965)n$ reaction in Feynman model and Regge model.}\label{KNh1uu}
\end{figure}

As we know, the $h_1(1965)$ as a further state is listed in PDG \cite{null}. Only the Crystal Barrel experiment \cite{Anisovich:2011sva} reported its resonance parameters by analyzing the $p\bar{p}\to \omega \eta$, $\omega\pi^0\pi^0$ reactions. It means that $h_1(1965)$ is still waiting for the confirmation from other experiments. Thus, the production of $h_1(1965)$ induced by pion and kaon on a proton target have also been studied in this work.
In Fig. \ref{piNh1uu}, we present our calculations of the total cross section for $\pi^-p\to h_1(1965)n$  in both the Feynman model and the Regge model. The line shape of total cross section in both models are similar to the case of $\pi^-p\to X(2100)n$. The total cross section reaches a maximum of 5.4 $\mu$b at ${\mathrm{p_{Lab}}}$=11.0 GeV$/c$ in the Regge model. In the Feynman case, the total cross section increase with enhancement of beam momentum, and then tend to be about 4 $\mu$b. We can see the production cross section of $h_1(1965)$ induced by pion is very significant, which is mainly benefited from a dominant decay width of $h_1(1965) \to \rho\pi$. In addition, the $h_1(1170)$ as ground state has been observed in pion-proton scattering \cite{Ando:1990ti}. And the measured cross section of the process $\pi^-p\to h_1(1170)n$ at laboratory momentum ${\mathrm{p_{Lab}}}=$ 8.06 GeV is $41.1\pm4.9$ $\mu$b as marked with a black star in Fig. \ref{piNh1uu}, which can be compared with our prediction. Thus, it is very promising to observe $h_1(1965)$ structure in pion-proton scattering experiment. Compared to pion induced reaction, the cross section of $K^-p\to h_1(1965)\Lambda$ presented in Fig. \ref{KNh1uu} is much smaller since $\mathcal{B}(h_1(1965)\to KK^*)$ is about two orders of magnitude smaller than $\mathcal{B}(h_1(1965)\to\pi\rho)$ and it is more difficult to exchange a $K^*$ meson than to exchange a $\rho$ meson. Therefore, we strongly recommend that the experiments with pion beam like J-PARC and COMPASS to study $h_1(1965)$, which can be detected by the dominant channel $\rho\pi$ and the reconstruction of final states $\pi^0\pi^+\pi^-n$ is suggested by us. Obviously the more deeper knowledge of $h_1(1965)$ is also helpful to understand the nature of $X(2100)$, so it is a good chance for the above experiment facilities.

\section{Summary}\label{sec4}

Inspired by the BESIII's observation of $X(2100)$ from analyzing $J/\psi\to \eta\eta^\prime \phi$ \cite{null}, we study possibility of $X(2100)$ as a conventional meson.
Through the analysis of the mass, we find that there exist two assignments to $X(2100)$, i.e., $X(2100)$ can be the second radial excitation of $h_1$ family or the second
radial excitation of isoscalar vector states.

Thus, a further calculation of two-body OZI-allowed strong decay behavior of $X(2100)$ under these possible assignments is helpful to test these assignments by combing with
the present experimental data. Our result shows that it is suitable to explain $X(2100)$ to be as the second radial excitation of $h_1(1380)$ since the experimental width
of $X(2100)$ can be reproduced and the branching ratio of $X(2100)\to\phi\eta^{\prime}$ has a sizable contribution to the total width. In addition, the $h_1(1965)$ is a good
candidate for the partner of the second radial excitation of $h_1(1380)$. We suggest a possible channel $J/\psi\to \phi\eta\eta$ to identify $X(2100)$ in the corresponding
$\phi\eta$ invariant mass spectrum. As an important part of light hadron spectrum, the axial-vector light meson family is not well established. The observation of $X(2100)$
in $J/\psi\to\eta\eta^\prime\phi$ provides us a good chance to further perform the investigation of isoscalar  axial-vector light mesons.

In addition, the productions of $X(2100)$ and its partner $h_1(1965)$ induced by pion and kaon on a proton target are predicted in both the Feynman model and the Regge model. The total cross sections are predicted to be at an order of magnitude of 0.1 $\mu$b in both $\pi^-p\to X(2100)n$ and $K^-p\to X(2100)\Lambda$. This result indicates $X(2100)$ is potential to be experimentally measured by J-PARC, COMPASS, OKA@U-70 and SPS@CERN experiment. The predicted cross section of $\pi^-p\to h_1(1965)n$ is much larger than that of the reaction $K^-p\to h_1(1965)\Lambda$ and can be compared with the experimental data of process $\pi^-p\to h_1(1170)n$. Therefore, we strongly recommend that COMPASS and other experiments to discover $h_1(1965)$ on the pion-proton scattering platform. This is of great significance for studying the properties of $h_1(1965)$ and understanding the newly observed $X(2100)$.

In the following several years, the BESIII and COMPASS experiments will still be the main force of exploring the light hadrons. These theoretical predictions presented in this work may provide valuable reference to future experimental studies on this issue.

In fact, the physics around 2.1 GeV light hadrons should be paid more attentions, which has close relation to these higher states of the $\rho$, $\phi$, $\omega$ and $h_1$ meson families, and exotic states.  The former observed $Y(2175)$ has inspired extensive discussion on this issue. When facing the different experimental observations of the states around 2.1 GeV, we should be very careful to directly treat them to be the same state as the observed $Y(2175)$ only according to a simple comparison of their resonance parameters. Furthermore, the study of their decay and production will provide valuable information to identify their inner structure. The present work provides a typical example. We expect more theoretical groups to focus on the physics around 2.1 GeV light hadrons.


\vfil
\section*{Acknowledgments}
This project is supported by the China National Funds for Distinguished Young Scientists under Grant No. 11825503
and the National Natural Science Foundation of China under Grant No. 11675228, and the National Program for Support of Top-Notch Young Professionals.


\end{document}